\documentclass{iaus}
\usepackage{graphicx}
\usepackage{multicol}

\title[Attenuation \& Photometric Redshifts] 
{The Impact of Stochastic Attenuation on Photometric Redshift Estimates}

\author[Tepper Garc\'\i a, Fritze-v.A.]   
{Thorsten Tepper Garc\'\i a$^1$%
  \break \and Uta Fritze-von Alvensleben$^2$}

\affiliation{$^1$Institut f\"ur Astrophysik, Georg-August Universit\"at, \break
Friedrich-Hund-Platz 1, 37077 G\"ottingen, Germany \break email:
tepper@astro.physik.uni-goettingen.de\\[\affilskip] $^2$Centre for Astrophysics
Research, University of Hertfordshire, \break College Lane, Hatfield AL10 9AB,
UK \break email: ufritze@star.herts.ac.uk}

\pubyear{2006}
\volume{235}  
\pagerange{1--}
\date{?? and in revised form ??}
\jname{Proceedings Galaxy Evolution across the Hubble Time}
\editors{F. Combes \& J. Palou\v s, eds.}

\newcommand{\ie}{
\textit{i.e.}
}

\newcommand{\ea}{
\textit{et al.}
}

\newcommand{\hi}{
H{\sc i}
}

\newcommand{\lya}{
Ly$\alpha$
}

\newcommand{\gv}{
GALEV
}

\begin{document}

\maketitle

\begin{abstract}
We model the stochastic attenuation by \hi{} absorbers in the intergalactic
medium (IGM), such as \lya{} Forest clouds, and absorbers associated with
galaxies, such as Lyman Limit systems (LLS) and Damped Lyman Alpha absorbers
(DLAs), and compute an ensemble of $4 \cdot 10^3$ attenuated Spectral Energy
Distributions (SEDs) in the Johnson system for the spectrum of a galaxy with a
constant star formation rate (CSFR). Using these, we asses the impact of the
stochastic attenuation on the estimates of photometric redshifts for this type of
galaxy by comparison with model SEDs that include only a mean attenuation.
\keywords{galaxies: intergalactic medium, galaxies: high-redshift, galaxies:
photometry} 
\end{abstract}

\firstsection
\section{Motivation}
The stochastic distribution of the (high column density) \hi absorbers causes a
significant scatter in the broadband colors of galaxies, especially at redshifts
close to the drop-out redshift in the corresponding passband where the
photoelectric absorption is more severe, and this is expected to drastically
affect the estimates of photometric redshifts.

\section{Approach}
We model the attenuation due to \hi along a random line of sight (LOS) using
differential distribution functions constrained from observations (\cite{Kim97})
in a Monte Carlo fashion (\cite{Bershady99}) as described in Tepper Garc\'\i a \&
Fritze-v.A. (in prep.). We generate an ensemble of $4 \cdot 10^{3}$ lines of
sight out to a given redshift $z_{em}$, each of them containing a random absorber
population. For each LOS we calculate an absorption mask, \ie we compute the
photoelectric and Lyman-Series line absorption (as yet just for the first five
Lyman transitions) caused by each absorber for a \textit{flat} input spectrum,
modeling line profiles as in \cite{Tepper06}. We compute model galaxy spectra
corresponding to a CSFR for redshifts in the range $0.0 < z_{em} < 4.5$ using the
Evolutionary Synthesis code \gv (\cite{Bicker04}). For a given redshift, we
multiply each of the $4\cdot10^3$ masks with our input spectrum and thus obtain
an ensemble of an equal number of attenuated spectra. For each of these we
compute a SED in the Johnson system. Using AnalySED (\cite{Anders06}) and a set 
of template SEDs that include only the mean attenuation (\cite{Madau95}) for
every redshift, we determine to which extent the redshifts of our simulated
spectral energy distributions are recovered.\\

\section{Results \& Conclusions}

As can be seen in figure 1, we find that, using the mean attenuation only, the
median photometric redshifts are severe underestimated by typically $\Delta
z_{phot} = 0.3$, especially in the range \mbox{$ z > 3.2$}. In addition, we
observe a substantial scatter of \mbox{$\Delta z_{phot} \approx \pm 0.5$}.

\begin{figure}\label{fig:zp}
\begin{center}
   \includegraphics[scale=0.4, angle=0]{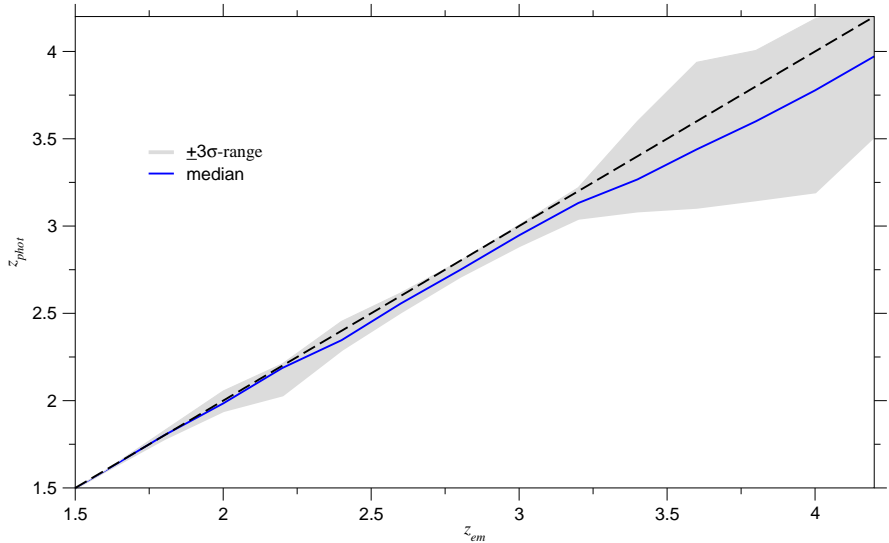}
   \caption{Estimate of the photometric redshift $z_{phot}$ for an Sd-type
   spectrum with \emph{stochastic} attenuation, using as input the same spectrum
   with \emph{mean} attenuation. The solid line shows the median, and the shaded
   area gives the $3\sigma$-range around the mean for the entire ensemble of
   4000 SEDs.}
\end{center}
\end{figure}

We predict that any method for determining photometric redshifts that only takes
into account the mean attenuation at high redshifts should show a bias such
as the one seen in Figure \ref{fig:zp}. We hence emphasise the need for an
accurate modeling of the attenuation including stochastic effects, in order to
correctly interpret in combination with evolutionary synthesis models such as
{\sc galev}, the observations of high-redshift galaxies in deep multi-band
imaging surveys.\\

\begin{acknowledgments}
TTG gratefully acknowledges support from the Georg-August-Universit\"at, from
CONACYT and the Astronomische Gesellschaft.
\end{acknowledgments}


\begin{thebibliography}{}

\bibitem[Anders \ea 2006]{Anders06}
	{Anders, P., Bissantz, N., Fritze-v. Alvensleben, U., de Grijs, R.} 2006,
	MNRAS, 347, 196

\bibitem[Bershady \ea 1999]{Bershady99}
	{Bershady, M.A., Charlton, J.C., Geoffroy, J.M.} 1999,
	ApJ, 518, 103
	
\bibitem[Bicker \ea 2004]{Bicker04}
	{Bicker, J., Fritze-v.A., U., Fricke, K.J.} 2004,
	A\&A, 413, 37

\bibitem[Kim \ea 1997]{Kim97}
	{Kim,  T.-S., Hu, E.M., Cowie, L.L., Songaila, A.} 1997,
	AJ, 114, 1

\bibitem[Madau 1995]{Madau95}
	{Madau, P.} 1995,
	ApJ, 441, 18

\bibitem[Tepper Garc\'\i a (2006)]{Tepper06}
	{Tepper Garc\'\i a, T.} 2006,
	MNRAS, 369, 2025

\end{thebibliography}
\end{document}